\documentclass[12pt]{iopart}

\begin{document}

\title{Far-infrared optical properties of the pyrochlore spin ice compound Dy%
$_2$Ti$_2$O$_7$}
\author{C.Z. Bi\dag\ddag\, J.Y. Ma\dag\, B.R. Zhao\dag\, Z. Tang\ddag\, D.
Yin\ddag\, C.Z. Li\ddag\, D.Z. Yao\ddag\, J. Shi\ddag\ and X.G. Qiu\dag\
\thanks{%
3}}
\address{\dag\ National Laboratory for Superconductivity, Institute of Physics,\\
Chinese Academy of Sciences, P.O. Box 603, Beijing 100080, China}

\address{\ddag\ Department of Physics, Wuhan University, Wuhan, Hubei
430072, China}
\begin{abstract}
Near normal incident far-infrared reflectivity spectra of [111]
dysprosium titanate (Dy$_2$Ti$_2$O$_7$) single crystal have been
measured at different temperatures. Seven phonon modes (eight at
low temperature) are identified at frequency below 1000 cm$^{-1}$.
Optical conductivity spectra are obtained by fitting all the
reflectivity spectra with the factorized form of the dielectric
function. Both the Born effective charges and the static optical
primitivity are found to increase with decreasing temperature.
Moreover, phonon linewidth narrowering and phonon modes shift with
decreasing temperature are also observed, which may result from
enhanced charge localization. The redshift of several low
frequency modes is attributed to the spin-phonon coupling. All
observed optical properties can be explained within the framework
of nearest neighbor ferromagnetic(FM) spin ice model.
\end{abstract}


\maketitle

\section{Introduction}

Recently, there has been a surge of interest in the properties of
pyrochlore compound Dy$_{2}$Ti$_{2}$O$_{7}$, which is considered
to be a model system of ``spin ice'' materials. ``Spin ice''
materials governed by the same statistical mechanics of so-called
``ice rule'' as the hydrogen atoms in the ground state of ordinary
hexagonal ice I$_{h}$ have macroscopically
degenerate ground states down to almost zero temperature.\cite%
{nature99,prb04,nature01,apa02, prb03,jpcm01,prl01,jpsj04,prb02}
Experimentally, the observed value of (1/2)Rln(3/2) through
specific heat measurement\cite{nature99} is consistent with what
is expected by pauling's theory, while the spin entropy only
freezes out below about 4 K.\cite{prb04} From magnetic
susceptibility studies, a strongly frequency dependent cooperative
spin freezing is observed at about 16 K, which is associated with
a very narrow distribution of spin relaxation times and a sharp
drop at about 2 K.\cite{nature01,jpcm01} Neutron scattering
studies performed by Fennell et al.\cite{apa02} also well
demonstrate the spin ice state and the coexistence of long range
ferromagnetic and short range antiferromagnetic
order in a magnetic field applied along the [110] axis of Dy$_{2}$Ti$_{2}$O$%
_{7}$. In theoretical aspect, R.G. Melko and coworkers\cite{prl01} report
numerical results on the low temperature properties of the dipolar spin ice
model by the multicanonical Monte Carlo (MC) method and they find a first
order transition to a long-range ordered phase. Other researchers\cite%
{jpsj04} also confirm the existence of the transition under a magnetic field
along the [110] axis with MC simulation.

Dy$_{2}$Ti$_{2}$O$_{7}$ has a typical A$_{2}$B$_{2}$O$_{7}$
structure with the space group (Fd3m, O$_{h}$$^{7}$), No.227. B
cation is sixfold coordinated and locates at the center of the
distorted octahedron formed by corner O ions. The A-site ions, i.e.,
Dy$^{3+}$(a magnetic rare-earth ion with effective spin S=1/2)
resides on a lattice of corner-sharing tetrahedral, as shown in
figure 1 of reference 3. In the Dy$^{3+}$ sublattice topology the
spin configuration with two spins pointing directly towards while
two spins pointing directly away from the center of the tetrahedral
corresponds to the analogous proton disorder in real ice. Then a
ferromagnetic and dipolar nearest-neighbor spin-spin interaction
leads to a strong geometrical frustration, preventing the system
from long range ordering. As a result, high degeneracy of spin
states will occur at low temperature.

Owing to the Ising anisotropy arising from crystal-field effect in
the pyrochlore lattice, the macroscopically degenerate states when
the field is along the [111] direction are different from those for
other field directions. For the [111] direction, the frustration
structure changes from that of a three-dimensional pyrochlore to
that of a two-dimensional Kagome-like lattice with constraint,
leading to different values of the zero-point entropy.\cite{prb03}
The combination of ferromagnetic coupling and Ising anisotropy may
be a reason for the strong frustration. But how the spins interact
with each other is not well understood in the spin ice materials.

Up to now, there has been no experimental report on the infrared optical
properties of Dy$_{2}$Ti$_{2}$O$_{7}$. IR spectroscopy can give insight to
the dynamical processes related to phonon, charge carrier, and spin.
Motivated by this situation, we studied the infrared reflectivity of Dy$_{2}$%
Ti$_{2}$O$_{7}$ single crystal on the [111] plane from room temperature down
to 10 K. The temperature dependence of phonon modes is obtained, and the
role played by spin-phonon coupling on the phonon modes is being discussed.

\section{Experimental details}

Dy$_{2}$Ti$_{2}$O$_{7}$ single crystal was prepared by the floating
zone method, using an infrared furnace equipped with two elliptical
mirrors. Before the single crystal growth, the polycrystalline rods
was prepared by a standard solid state reaction. Stoichiometric
mixture of Dy$_{2}$O$_{3}$
(CERAC, 99.99\%) and TiO$_{2}$ (CERAC, 99.9\%) was heated in air at 1250$%
^{o} $C for five days with intermediate regrinding to ensure
complete reaction. The typical growth condition was 4.5 mm/h for
both feed and growth speeds. To avoid oxygen deficiency, the single
crystal was grown in an O$_{2} $ atmosphere of 0.3 MPa. The single
crystal obtained was translucent yellow. Powder x-ray diffraction
measurements on the crystal confirmed that the product was a single
phase with cubic pyrochlore structure and the parameter of the unit
cell was a=10.1122\AA . The principal axes were determined using
Laue diffraction pattern.

The crystal was cleaved and one-side polished with the surface
parallel to the [111] plane. The reflectivity spectra R($\omega$)
at different temperatures were measured at near-normal incidence
of about 8$^o$ on a Bomen DA8 Fourier transform infrared
spectrometer, in the range from 50 to 6500 cm$^{-1}$. In the far
infrared region, a He cooled bolometer and 6 mm mylar beamsplitter
were used. In the middle infrared region, a liquid nitrogen cooled
mercury cadmium telluride detector and KBr beamsplitter were
utilized. To obtain the absolute reflectivity, an evaporated
golden mirror
was served as a reference. Spectra were collected with a resolution of 4 cm$%
^{-1}$. The samples were mounted in a continuous helium flow cryostat in
which the temperature could be varied between 300 and 10 K.

\section{Results and discussion}

The temperature dependent reflectivity of the
Dy$_{2}$Ti$_{2}$O$_{7}$ single crystal shown in figure 1 is typical
that of a nonmetallic system. The sharp features in the reflectivity
spectra are due to the unscreened infrared active optical phonon
modes, above the highest observed lattice vibration frequency, the
reflectivity is flat and featureless up to the highest measured
frequency. It can be seen that the reflectivity within the
reststrahlen bands increases with decreasing temperature. At high
temperature seven modes can easily be identified.

The mode frequencies are obtained from a damped harmonic oscillator
fit of reflectivity spectra with a complex dielectric function
$\varepsilon (\omega
) $ in the factorized form (generalized Lyddane-Sachs-Teller relation) \cite%
{infrared,prb95}:
\begin{eqnarray}
{\varepsilon (\omega )}=\varepsilon _{1}(\omega )+\varepsilon
_{2}(\omega )=\varepsilon _{\infty }\prod_{j=1}^{n}\frac{\Omega
_{jLO}^{2}-\omega ^{2}-\rmi\gamma _{jLO}\omega }{\Omega
_{jTO}^{2}-\omega ^{2}-\rmi\gamma _{jTO}\omega },
\end{eqnarray}
where $\varepsilon _{\infty }$ represents the high frequency
dielectric constant (at frequencies large compared with the
lattice vibration
frequencies but small compared with the electronic transition frequencies), $%
\Omega _{jLO}$ and $\Omega _{jTO}$ are the longitudinal and
transverse eigenfrequencies of the jth optical phonon mode, $\gamma
_{jLO}$ and $\gamma _{jTO}$ represent their longitudinal and
transverse damping constants. Using this dielectric function, all
the reflectivity spectra in our measurements were fitted with the
well known Fresnel formula for reflectivity of a half space medium
in vacuum\cite{prb91}:
\begin{eqnarray}
{R(\omega )}=\left\vert \frac{\sqrt{\varepsilon (\omega )}-1}{\sqrt{%
\varepsilon (\omega )}+1}\right\vert ^{2}
\end{eqnarray}
Based on equation (1) and (2), a fit of R(w) to the observed
reflectivity spectra
can be obtained with a proper choice of the model parameters $\Omega _{jLO}$%
, $\Omega _{jTO}$, $\gamma _{jLO}$, $\gamma _{jTO}$ and $\varepsilon
_{\infty }$ . Adjustment of the parameters is made by trial and error
fitting of formula (2) to the experimental spectra. This method yields not
only $\varepsilon (\omega )$ but also the model parameters which
characterize the infrared active phonons.

From a group theoretical analysis,\cite{infraram} a compound with
the pyrochlore structure symmetry should have the vibrational
phonon modes at the Brillouin zone center $\Gamma $ point
with\newline $\Gamma$=8 F$_{1u}$+4 F$_{2u}$+2 F$_{1g}$ +4 F$_{2g}$
+3 E$_{u}$ + E$_{g}$ + A$_{1g}$ +3 A$_{2u}$.\newline Among these
26 normal modes, only A$_{1g}$, E$_{g}$, 4 F$_{2g}$ are Raman
active, 7F$_{1u}$ are infrared active, and one F$_{1u}$
acoustical. Consequently, the observed phonons in the spectra of
Dy$_{2}$Ti$_{2}$O$_{7}$ at low temperature except that at about
610 cm$^{-1}$ were denoted
sequentially from low to high frequencies by F$_{1u}$$^{1}$ to F$_{1u}$$^{7}$%
. The number is in good agreement with the group theoretical
analysis. The mode at 610 cm$^{-1}$, denoted by F$_{1u}$$^{7\ast }$,
is weak in intensity and does not change with the temperature. It is
probably originated from two phonon absorption because of its low
intensity.\cite{pss98}

The fitting results of the oscillator parameters for all the phonon modes of
Dy$_{2}$Ti$_{2}$O$_{7}$ are summarized in Table~\ref{tab:table1}. The high
frequency dielectric constant is adopted as 5.0. Figure 2 shows the
experimental and fitting curves at two representative temperatures of 300 K
and 10 K. The fitting results with the Lorentz oscillators could explain the
global features of the phonon spectra reasonably well, even though some
difference between the experimental and the fitted data remains. A small
deviation of the calculated curves from the experimental one is seen on the
low frequency edge of F$_{1u}$$^{2}$ mode in the high temperature. A
possible reason is the reflection from the rear surface of the sample due to
its transparency at low frequencies as well as the cutoff of the working
frequency of the detector. However, the deviation has few impact on other
phonon fitting parameters in other frequency range.

The real part of the optical conductivity $\sigma _{1}(\omega )$
can be extracted from $\varepsilon _{2}(\omega )$, where $\sigma
_{1}(\omega )$=$1.6638\cdot10^{-2}\omega\varepsilon _{2}(\omega
)$, here $\omega$ is in unit of cm$^{-1}$ and $\sigma _{1}$ in
unit of $\Omega ^{-1}$ cm$^{-1}$, the corresponding result is
shown in figure 3.\cite{unit} The conductivity is dominated by
seven peak structures due to optical phonon absorption, no free
carrier contribution can be observed. For normal materials, with
decreasing T, the anharmonic thermal motion would decrease, which
results in a decrease in the lattice constant. Then the phonons
should shift to higher frequencies and their linewidth should
become narrower. The F$_{1u}$$^{6}$ mode has clearly demonstrated
this effect. Its center frequency shifts from 439 cm$^{-1}$ to 452
cm$^{-1}$ as the temperature decreases from 300 K down to 10 K.
Moreover, the phonon peak is more and more prominent and separated
from the nearby mode F$_{1u}$$^{5}$ . But it is puzzling that
other modes do not exhibit the same behavior, the
three most prominent phonons, F$_{1u}$$^{2}$, F$_{1u}$$^{3}$ and F$_{1u}$$%
^{5}$, show discernible redshifts. Coupling of phonons to magnetic
excitations in solids may result in a change of phonon self-energy, i.e.,
the frequency due to spin-phonon interaction. Therefore, spin-phonon coupling%
\cite{prb04lee} closely correlated with the Dy$^{3+}$ (its effective spin
S=1/2) could be a possible origin for the abnormal temperature dependence of
the phonon frequency. As we mentioned in the introduction, Dy$^{3+}$ cations
can have appreciable magnetic moments. Consequently, ferromagnetic nearest
neighbor dipole-dipole interactions will present in spin ice Dy$_{2}$Ti$_{2}$O$%
_{7}$ in which large dipole interactions have been suggested to be
responsible for the spin ice behavior.\cite{prl00} These effects have been
quantitatively confirmed by both experimental and theoretical approaches.%
\cite{prb02} At the nearest neighboring sites, the exchange interactions
between the magnetic atoms become the strongest. Subsequently, the crystal
potential U can be described as following:
\begin{eqnarray}
{U}=\frac{1}{2}kx^{2}+\sum_{ij}J_{ij}<S_{i}S_{j}>,
\end{eqnarray}
where x is the atomic displacement from the equilibrium position
in the oscillator model. In the second term, J$_{ij}$ is the
exchange energy constant which is a function of the structural
parameters, such as the bond lengths between Dy$^{3+}$cations and
the corresponding bond angles mediated by the O ions. S$_{i}$
represents the spin of Dy$^{3+}$cation at the ith site.
$<S_{i}S_{j}>$ denotes a statistical average over the adjacent
spins. The harmonic force constant derived from its second
derivative formula reads:
\begin{eqnarray}
{\frac{\partial ^{2}U}{\partial x^{2}}}=k+\sum_{ij}(\frac{\partial ^{2}J_{ij}%
}{\partial x^{2}})<S_{i}S_{j}>.
\end{eqnarray}
Note that the second term represents the spin-phonon coupling,
which suggests that the phonon frequency should have an additional
contribution. As for the spin-phonon coupling coefficient,
$\sum_{ij}(\frac{\partial ^{2}J_{ij}}{\partial x^{2}})$ can be
different for each phonon and can have either a positive or a
negative sign. Furthermore, different phonon frequencies will have
redshift or blueshift in the optical conductivity spectra. As the
temperature decreases, spin fluctuation becomes weaker and
spin-phonon coupling stronger. It is important to notice that
although ferromagnetic correlations are short ranged without long
range order in spin ice compounds, from neutron scattering
measurements Harris et al.\cite{prl97} have shown that the range
of ferromagnetic order increases at lower temperature. As a
result, spin-phonon coupling range is wider so that the phonon
frequency shows redshift or blueshift, which is in agreement with
what is observed in our optical conductivity spectra.

Now we focus on the remarkable change of the spectra weight which is
proportional to the area under the optical conductivity peak. It is
well known that, by decreasing the temperature, thermal fluctuation
reduces, which may result in narrowing in the linewidth and
enhancement in the oscillator strength without any changes in the
bonding or coordination. Almost all phonon modes exhibit the
theoretically predicted behavior in our optical conductivity
spectra. But the spectral weight should not change. The anomalous
increase in oscillator strength of the low-frequency modes must be
taken sufficient care of. Optical sum rules provide a powerful tool
with
which to analyze the behavior of free carriers and bound excitations.\cite%
{handbook} The partial conductivity sum rules corresponding
primarily to a single class of absorption such as excitation of
phonons, conduction, valence, or core electrons have been developed.
For oscillator states, the partial conductivity sum rule can be
expressed as \cite{prb03homes}
\begin{eqnarray}
{\frac{120}{\pi }\int_{\omega _{a}}^{\omega _{b}}\sigma
_{1}(\omega )d\omega }=\omega _{p,j}^{2},
\end{eqnarray}
where $\omega _{a}$, $\omega _{b}$ and $\omega _{p,j}$ (in unit of cm$^{-1}$%
) are the integral lower and upper limits and effective plasma
frequency associated with the isolated phonon absorption,
respectively. In the above equation, $\sigma $ is in unit of $\Omega
^{-1}cm^{-1}$. The integral region of the jth oscillator should be
chosen to cover the full spectral weight. The dramatic increase in
the spectra weight has implications for the distribution of charge
and the change in local strength of the binding charge. The
effective charge of Dy, Ti and O ions in the unit cell with k atoms
can be determined through the following Equation (6)(in Gauss unit
system)\cite{prb71}
\begin{eqnarray}
{\frac{1}{\varepsilon _{\infty }}\sum_{j}\omega _{p,j}^{2}}=\frac{4\pi }{%
V_{c}}\sum_{k}\frac{(Z_{k}^{\ast }e)^{2}}{M_{k}},
\end{eqnarray}
where V$_{c}$ is the unit cell volume, j and k index the lattice
modes and the atoms with mass M$_{k}$, respectively. For the
effective charge, there is a general expression
$\sum_{k}Z_{k}^{\ast }$=0.

In Dy$_{2}$Ti$_{2}$O$_{7}$, oxygen is the lightest element, therefore, in
the right side of equation (6), the summation is dominated by O ion item and
the terms for Dy ion and Ti ion may be neglected. The change in the
effective charge is associated mainly with the oxygen (i.e., $Z_{k}^{\ast }$$%
\approx $$Z_{o}^{\ast }$). Combining equation (5) and equation (6),
we obtain the value for $Z_{o}^{\ast }$ as shown in figure 4. The
absolute value of $Z_{o}^{\ast }$ is of less significance, what is
important is the temperature dependence of the deduced value for
$Z_{o}^{\ast }$, which increases with decreasing temperature. When
the incident light couples to the induced dipole moments created by
the atomic displacements associated
with a normal mode, if the Born effective charge per oxygen atom $%
Z_{o}^{\ast }$ is increasing, then the size of the induced dipole
moment and the optical absorption will also increase, which well
explain the above temperature dependence of the optical
conductivity. On the other hand, the increase in $Z_{o}^{\ast }$
means a change in the bond length and bond angle between O ion and
cations so that the bond lengths between Dy$^{3+}$ cations and the
corresponding bond angles mediated by the O ions will change.
Subsequently, J$_{ij}$ and $<S_{i}S_{j}>$ have different changes
associated with phonon central frequency shift as we observed
above. In summary, the change in the effective charge arising from
electrical charge localization strengthening is the fundamental
reason why the spectra weight increases and phonon frequency
shifts. From our experimental results, within the framework
of the ferromagnetic spin ice model, we consider that the increase in $%
Z_{o}^{\ast }$ is closely related with nearest-neighbor ferromagnetic
interaction strengthening and/or widening in the correlation range. In this
simple physical picture, the increasing FM exchange between Dy$^{3+}$
cations will lead to a decrease of the distance between mediated O ion and Dy%
$^{3+}$, which enhance the increase in $Z_{o}^{\ast }$, in agreement with
our experimental observation.

LST relationship also provides an effective examination on static optical
permittivities $\varepsilon _{0}$ in the approximation of zero phonon
frequency. When $\omega $=0 , the equation evolutes into the following form:
\begin{eqnarray}
{\frac{\varepsilon _{0}}{\varepsilon _{\infty }}}=\prod_{i}\frac{\Omega
_{iLO}^{2}}{\Omega _{iTO}^{2}}.
\end{eqnarray}
From Equation (7) $\varepsilon _{0}$ of Dy$_{2}$Ti$_{2}$O$_{7}$
single crystal at different temperatures are also obtained, as
shown in the inset of figure 4. It exhibits the same temperature
dependence as $Z_{o}^{\ast }$, which suggests that the temperature
dependence of $\varepsilon _{0}$ can be originated from that of
$Z_{o}^{\ast }$ and essentially, charge localization arising from
FM exchange plays an important role in the temperature variation
of both $\varepsilon _{0}$ and $Z_{o}^{\ast }$.

\section{Conclusions}

In conclusion, FIR response of Dy$_{2}$Ti$_{2}$O$_{7}$ single
crystals has been studied under different temperatures. Seven
infrared active phonons have been observed, which agrees well with a
group analysis. All the spectra are fitted with the oscillator model
and an excellent agreement between experimental and calculated
spectra is obtained. Of all phonons, one shows discernible blueshift
and three obvious redshifts with decreasing temperature. The shifts
are attributed to the spin-phonon coupling in a geometrically
frustrated configuration in the spin ice material. The oscillation
strength of the low frequency modes increase dramatically at low
temperature, indicating that the Born effective charges are
increasing in the unit cell. We propose that the similar temperature
dependence of the static optical permittivity with Born effective
charge may originate from the intrinsic charge localization resulted
from the nearest-neighbor FM interaction.

\ack
This work is supported by National Science Foundation of China (Grants No.
10474128 and No. 10474074).

\newpage Figure captions

Figure 1 The temperature dependence of the reflectance of Dy$_{2}$Ti$_{2}$O$%
_{7} $ single crystal in the range of ~50-1000 cm$^{-1}$.

Figure 2 Representative experimental and fitted reflectivity spectra of Dy$%
_2 $Ti$_2$O$_7$ at (a) 300 K and (b) 10 K. Solid lines represent
experimental data and dot lines are the fitting results.

Figure 3 The real part of the temperature dependent optical
conductivity of Dy$_2$Ti$_2$O$_7$ in the far-infrared region.

Figure 4 The temperature dependence of Born effective charge per
oxygen atom in Dy$_2$Ti$_2$O$_7$. The dots represent the deduced
values at various temperatures between 300 and 10 K. Inset: the
temperature dependence of the static optical permittivities.

\newpage
\begin{table}[tbp]
\caption{The phonon parameters for the Lorentzian fits to the conductivity
of Dy$_2$Ti$_2$O$_7$ single crystal at different temperatures. All units are
in cm$^{-1}$.}
\label{tab:table1}%
\begin{indented}
\item[]\begin{tabular}{lccccccccc}
\br
 $Temperature$&&&&&$F_{1u}$&$modes$&\\
\mr
 T=300 K& $\Omega_{TO}$&$-$& 137 &226
& 261 & 374 & 440 & 545 & 608 \\
$ $&$\gamma_{TO}$ &$-$& 23 &77
& 44 & 62 & 26 & 36 & 18 \\
$ $&$\Omega_{LO}$ &$-$& 143 &253
& 320 & 437 & 537 & 608 & 745 \\
$ $&$\gamma_{LO}$ &$-$& 12 &31
& 33 & 26 & 38 & 19 & 36 \\
T=250 K&$\Omega_{TO}$ &$-$& 133 &224
& 261 & 372 & 442 & 544 & 611 \\
$ $&$\gamma_{TO}$ &$-$& 18 &65
& 40 & 60 & 22 & 31 & 18 \\
$ $&$\Omega_{LO}$ &$-$& 142 &253
& 320 & 440 & 537 & 611 & 742 \\
$ $&$\gamma_{LO}$ &$-$& 13 &29
& 31 & 22 & 33 & 19 & 26 \\
T=200 K& $\Omega_{TO}$&93& 135 &204
& 259 & 370 & 444.5 & 543 & 608 \\
$ $&$\gamma_{TO}$ &15& 23 &62
& 26 & 42 & 16 & 23 & 17 \\
$ $&$\Omega_{LO}$ &93.5& 144 &254
& 322 & 443 & 538 & 608 & 746 \\
$ $&$\gamma_{LO}$ &12& 12 &22
& 29 & 16 & 24 & 18 & 20 \\
T=150 K& $\Omega_{TO}$&91.5& 130 &199
& 259 & 370 & 447 & 544 & 608 \\
$ $&$\gamma_{TO}$ &18& 18 &45
& 45 & 35 & 16 & 23 & 16.5 \\
$ $&$\Omega_{LO}$ &92& 142 &254
& 321 & 445 & 538 & 608 & 749 \\
$ $&$\gamma_{LO}$ &14& 13 &20
& 29 & 16 & 25 & 17 & 28 \\
T=100 K& $\Omega_{TO}$&88& 127 &198
& 258 & 370 & 450 & 546 & 607 \\
$ $&$\gamma_{TO}$ &18& 12 &36
& 20 & 33 & 19 & 29 & 20 \\
$ $&$\Omega_{LO}$ &90& 140 &252
& 320 & 447.5 & 537 & 607 & 748 \\
$ $&$\gamma_{LO}$ &23& 12 &20
& 30 & 20 & 35 & 21 & 45 \\
T=50 K& $\Omega_{TO}$&86& 127 &197
& 259 & 370 & 453 & 545 & 612 \\
$ $&$\gamma_{TO}$ &15& 10 &29
& 18 & 31 & 19 & 30 & 20 \\
$ $&$\Omega_{LO}$ &89& 141 &252
& 320 & 450 & 536 & 612 & 748 \\
$ $&$\gamma_{LO}$ &29& 15 &18
& 28 & 20 & 34 & 21 & 45 \\
T=10 K& $\Omega_{TO}$&85& 126 &197
& 260 & 370 & 454 & 545 & 612 \\
$ $&$\gamma_{TO}$ &15& 7 &26
& 17 & 27 & 18 & 31 & 22 \\
$ $&$\Omega_{LO}$ &90& 139 &253
& 317 & 451 & 535 & 612 & 748 \\
$ $&$\gamma_{LO}$ &27& 17 &18
& 26 & 20 & 34 & 23 & 43 \\
\br
\end{tabular}
\end{indented}
\end{table}


\section*{References}

\begin{thebibliography}{99}
\bibitem{nature99} Ramirez A P, Hayashi A, Cava R J, Siddharthan R and
Shastry B S 1999 \textit{Nature (London)} \textbf{399} 333

\bibitem{prb04} Snyder J, Ueland B G, Slusky J S, Karunadasa H, Cava R J and
Schiffer P 2004 \textit{Phys. Rev.} B \textbf{69} 064414

\bibitem{nature01} Snyder J, Slusky J S, Cava R J and Schiffer P 2001
\textit{Nature (London)} \textbf{413} 48

\bibitem{apa02} Fennell T, Petrenko O A, Balakrishnan G, Bramwell S T,
Champion J D M, Fak B, Harris M J and Paul D M 2002 \textit{Appl. Phys.} A
\textbf{74} S889

\bibitem{prb03} Higashinaka R, Fukazawa H and Maeno Y 2003 \textit{Phys. Rev.%
} B \textbf{68} 014415

\bibitem{jpcm01} Matsuhira K, Hinatsu Y and Sakakibara T 2001 \textit{J.
Phys.: Condens.Matter} \textbf{13} L737

\bibitem{prl01} Melko R G, den Hertog B C and Gingras M J P 2001 \textit{%
Phys. Rev. Lett.} \textbf{87} 067203

\bibitem{jpsj04} Yoshida S, Nemoto K and Wada K 2004 \textit{J. Phys. Soc.
Jpn.} \textbf{71} 1619

\bibitem{prb02} Fukazawa H, Melko R G, Higashinaka R, Maeno Y and Gingras M
2002 \textit{Phys. Rev.} B \textbf{65} 054410

\bibitem{infrared} Gervais F 1983 \textit{in Infrared and Millimetre Waves}
Vol. 8 edited by Button K J (Academic, New York) p. 279

\bibitem{prb95} Massa N E, Campa J and Rasines I 1995 \textit{Phys. Rev.} B
\textbf{52} 15920

\bibitem{prb91} Tajima S, Ido T, Ishibashi S, Itoh T, Eisaki H, Mizuo Y,
Arima T, Takagi H and Uchida S 1991 \textit{Phys. Rev.} B \textbf{43} 10496

\bibitem{infraram} Fateley W G, Dollish F R, McDevitt N T and Bentley F F
1972 \textit{Infrared and Raman Selection Rules for Molecular and
LatticeVibrations: The Correlation Method} (Wiley-Interscience, NewYork)

\bibitem{pss98} Kamba S, Buixaderas E and Pajaczkowska A 1998 \textit{Phys.
Status Solidi} A \textbf{168} 317

\bibitem{unit} Bruesch P 1986 \textit{Phonons: Theory and Experiments
II} (Springer-Verlag, Berlin) Ch. 2 p. 14

\bibitem{prb04lee} Lee J S, Noh T W, Bae J S, Yang In-San , Takeda T and
Kanno R 2004 \textit{Phys. Rev.} B \textbf{69} 214428

\bibitem{prl00} den Hertog B C and Gingras M J P 2000 \textit{Phys. Rev. Lett%
} \textbf{84} 3430

\bibitem{prl97} Harris M J, Bramwell S T, McMorow D F, Zeiske T and Godfrey
K W 2000 \textit{Phys. Rev. Lett} \textbf{79} 2554

\bibitem{handbook} Smith D Y 1985 \textit{Handbook of Optical Constants of
Solids} (Academic, New York).

\bibitem{prb03homes} Homes C C, Vogt T, Shapiro S M, Wakimoto S, Subramanian
M A and Ramirez A P 2003 \textit{Phys. Rev.} B \textbf{67} 092106

\bibitem{prb71} Scott J F 1997 \textit{Phys. Rev.} B \textbf{4} 1360
\end{thebibliography}
\end{document}